\documentstyle[11pt]{article}

\newlength{\dinwidth}
\newlength{\dinmargin}
\def\slepton{\widetilde \ell}
\def\sl{{\widetilde \ell}^{-}}
\def\slb{{\widetilde \ell}^{+}}
\def\slr{\slepton_{R}}

\def\squark{\widetilde q}
\def\msl{m_{\slepton}}
\def\mslr{m_{\slr}}

\def\sele{\widetilde e}
\def\ser{\sele_{R}}

\def\smu{\widetilde \mu}
\def\smr{\smu_{R}}

\def\stau{\widetilde \tau}

\def\su{\widetilde{u}}
\def\sd{\widetilde{d}}
\def\st{\widetilde{t}}

\def\ddf{{\rm d}}
\def\bino{\widetilde{B}}

\def\sz1{{\widetilde{Z}}_{1}}

\def\msz1{m_{\sz1}}

\def\mbino{m_{\bino}}

\def\rs{{\sqrt{s}}}

\def\nle{{\stackrel{<}{\sim}}}
\def\nge{{\stackrel{>}{\sim}}}

\def\sh{{\hat{s}}}
\def\bh{{\hat{\beta}}}
\def\bhd{{\hat{\beta'}}}
\def\rsh{\sqrt{\sh}}

\begin{document}
{}~~~\\
% \vspace{10mm}
\begin{flushright}
ITP-SU-93/04 \\
\end{flushright}
\begin{center}
  \begin{Large}
  \begin{bf}
\renewcommand{\thefootnote}{\fnsymbol{footnote}}
Polarization Effects in Scalar Lepton Production \\
at High Energy $\gamma\gamma$ Colliders
 \\
  \end{bf}
  \end{Large}
  \vspace{5mm}
  \begin{large}
\renewcommand{\thefootnote}{\fnsymbol{footnote}}
    Tadashi Kon
\footnote{
e-mail address : d34477@jpnac.bitnet, mtsk@jpnkektr.bitnet
}
\\
  \end{large}
Faculty of Engineering, Seikei University, Tokyo 180, Japan
  \vspace{5mm}
\end{center}
\begin{quotation}
\noindent
{\bf Abstract:}
We investigate the charged scalar lepton production processes
$\gamma\gamma\to
{\widetilde{\ell}}^{+}{\widetilde{\ell}}^{-}$
at high energy $\gamma\gamma$ colliders
in the framework of the minimal supersymmetric standard model
(MSSM).
Here the high energy $\gamma$ beams are obtained by the backward
Compton scattering of the laser flush by the electron in the
basic linear TeV $ee$ colliders.
We consider the polarization of the laser photons as well as the
electron beams.
Appropriate beam polarization could be effective to
enhance the cross section and to extract the signal from
the dominant background $\gamma\gamma\to W^{+}W^{-}$.

\end{quotation}
%
%\section{Introduction}

In addition to well-known types of "TeV" colliders such as
$pp$, $ep$ and linear $e^{+}e^{-}$ colliders,
the possibilities for realization of the
$e\gamma$ and $\gamma\gamma$ colliders have been discussed in
detail by Ginzburg et al. \cite{Ginz}.
Here the high energy photon beams will be obtained by the
backward Compton scattering of the laser flush by one of
electron beam in the basic linear $ee$ colliders.
Recently, the physics potential at those colliders has been
analyzed energetically from both the theoretical \cite{Brodsky}
and experimental \cite{Borden} point of view.
In fact such colliders will provide us with a powerful
machinery for investigating the standard model (SM) through a direct
proof of the gauge vertices \cite{gauge}, unraveling
the hadronic content of the photon \cite{QCD} and
searching for the neutral Higgs boson \cite{Higgs}.
The $e\gamma$ and $\gamma\gamma$ colliders are
also suited for searching for
some exotic particles predicted by the models beyond the standard model,
such as supersymmetric partners \cite{egamma,gamma,Cuypers,proc},
excited fermions \cite{GI,egest} and the leptoquarks \cite{Eboli}.

In this paper we investigate
the charged scalar lepton (slepton) production
at the $\gamma\gamma$ colliders.
The sparticle production at the $\gamma\gamma$ colliders,
including the slepton production, has already been discussed in
Ref.\cite{gamma,Cuypers}.
In those papers, however, the initial beam polarization has
not been concerned.
Here we focus our attention to
the physical consequence of the initial beam polarization
at the $\gamma\gamma$ colliders
in the slepton production.

%\section{Calculation}

The polarized cross sections for
the sub-processes $\gamma\gamma\to X$ are expressed as
\begin{eqnarray}
&&\ddf{\hat{\sigma}}({\xi_{2}(z_{1}),\xi_{2}(z_{2})}) \nonumber \\
&&={\frac{1}{4}}\biggl[
\left(1+\xi_{2}(z_{1})\right)\left(1+\xi_{2}(z_{2})\right)
\ddf{\hat{\sigma}}[\gamma_{+}\gamma_{+}]+
\left(1+\xi_{2}(z_{1})\right)\left(1-\xi_{2}(z_{2})\right)
\ddf{\hat{\sigma}}[\gamma_{+}\gamma_{-}]+ \nonumber \\
&& \qquad \left(1-\xi_{2}(z_{1})\right)\left(1+\xi_{2}(z_{2})\right)
\ddf{\hat{\sigma}}[\gamma_{-}\gamma_{+}]+
\left(1-\xi_{2}(z_{1})\right)\left(1-\xi_{2}(z_{2})\right)
\ddf{\hat{\sigma}}[\gamma_{-}\gamma_{-}]\biggr],
\label{polsub}
\end{eqnarray}
where $\xi_{2}(z_{i})$ denote
the Stokes parameters of the circular polarized $\gamma$ beams.
Note that there are no terms proportional to
$\xi_{1}$ and $\xi_{3}$ because we have assumed that the initial
laser lights are circularly polarized and thus the secondary
gamma beams are also circulaly polarized
($\xi_{1}$ $=$ $\xi_{3}$ $=$ 0).

The each polarized cross sections in Eq.(\ref{polsub})
for the slepton pair production
$\gamma\gamma\to \slb\sl$ are given by
\begin{eqnarray}
&&{\frac{\ddf{\hat{\sigma}}}{\ddf\cos\theta}}
  [\gamma_{\pm}\gamma_{\pm}\to\slb\sl]=
  {\frac{2\pi\alpha^{2}}{\sh(1-\bh^{2}\cos^{2}\theta)^{2}}}
  \bh(1-\bh^{2})^{2}\nonumber \\
&&{\frac{\ddf{\hat{\sigma}}}{\ddf\cos\theta}}
  [\gamma_{\pm}\gamma_{\mp}\to\slb\sl]=
  {\frac{2\pi\alpha^{2}}{\sh(1-\bh^{2}\cos^{2}\theta)^{2}}}
  \bh^{5}(1-\cos^{2}\theta)^{2},
\label{subsl}
\end{eqnarray}
where $\sh\equiv s_{\gamma\gamma}$ and $\bh\equiv\sqrt{1-4{\msl}^{2}/\sh}$.
The total cross sections are given by
\begin{eqnarray}
&&{\hat{\sigma}}
  [\gamma_{\pm}\gamma_{\pm}\to\slb\sl]=
  {\frac{2\pi\alpha^{2}}{\sh}}
  \bh(1-\bh^{2})\left[1+{\frac{1-\bh^{2}}{2\bh}}
\ln{\frac{1+\bh^{2}}{1-\bh^{2}}}\right]\nonumber \\
&&{\hat{\sigma}}
  [\gamma_{\pm}\gamma_{\mp}\to\slb\sl]=
  {\frac{2\pi\alpha^{2}}{\sh}}
  \bh\left[3-\bh^{2}+{\frac{1}{2\bh}}(-3+2\bh^{2}+\bh^{4})
\ln{\frac{1+\bh^{2}}{1-\bh^{2}}}\right].
\label{total}
\end{eqnarray}
Since our processes are pure SUSY QED processes,
we can express cross sections for all charged sleptons
$\slepton =$ ($\sele_{L,R}$, $\smu_{L,R}$, $\stau_{1,2}$)
by the formulae Eqs.(\ref{subsl}) and (\ref{total}).
Moreover, multiplying the factor $e_{q}^{4}C$
($e_{q}$ and $C=3$ respectively denote the electric charge
and the color degree of freedom),
we can use the formulae Eqs.(\ref{subsl}) and (\ref{total}) for the squark
$\squark =$ ($\su_{L,R}$, $\sd_{L,R}$, $\cdots$, $\st_{1,2}$)
pair production.
For completeness, we also give the formula for the sub-process
cross section of $\gamma\gamma\to W^{+}W^{-}$,
which will be needed in the discussion of background suppression
;
\begin{eqnarray}
&&{\frac{\ddf{\hat{\sigma}}}{\ddf\cos\theta}}
  [\gamma_{\pm}\gamma_{\pm}\to W^{+}W^{-}]=
  {\frac{2\pi\alpha^{2}}{\sh(1-\bhd^{2}\cos^{2}\theta)^{2}}}
  \bhd(3+10\bhd^{2}+3\bhd^{4})\nonumber \\
&&{\frac{\ddf{\hat{\sigma}}}{\ddf\cos\theta}}
  [\gamma_{\pm}\gamma_{\mp}\to W^{+}W^{-}] = \nonumber \\
&& {\frac{2\pi\alpha^{2}}{\sh(1-\bhd^{2}\cos^{2}\theta)^{2}}}
  \bhd\left[16-16\bhd^{2}+3\bhd^{4}
 +2\bhd^{2}(8-3\bhd^{2})\cos^{2}\theta
+3\bhd^{4}\cos^{4}\theta\right],
\label{subW}
\end{eqnarray}
where $\bhd\equiv\sqrt{1-4m_{W}^{2}/\sh}$.
Our results Eq.(\ref{subW}) are in agreement with the ones of
Ginzburg et al. \cite{gauge}.

If the incident gamma beams were monochromatic,
the formulae (\ref{polsub}),(\ref{subsl}),(\ref{total}) and (\ref{subW})
would give just the cross sections.
However, since each gamma is the secondary beam, the experimental
cross sections are obtained by folding the sub-process
cross sections
$\ddf\hat{\sigma}$ (\ref{subsl}) and (\ref{subW}) with
the energy spectra $D_{\gamma/e}$
of the high energy photons generated by
the backward Compton scattering of the laser light ;
\begin{equation}
\ddf\sigma=
\int_{0}^{z_{1}^{max}}\ddf z_{1}\int_{0}^{z_{2}^{max}}\ddf z_{2}
D_{\gamma/e}(z_{1})D_{\gamma/e}(z_{2})
\ \ddf{\hat{\sigma}} \
\theta(z_{1}z_{2}s_{ee}-4\msl^{2}).
\label{expsig}
\end{equation}
Here $z_{i}$ ($i=1,2$) denotes the energy fraction of
each high energy gamma beam ;
$z_{i}=E_{\gamma_{i}}/E_{e}$. Note that the upper
limit of $z_{i}$ is determined by the kinematics
of the backward-Compton scattering, i.e., $z_{i}\nle 0.83$.
If $z_{i}$ becomes larger than this value the back-scattered and
laser photon have enough energy to produce the $e^{+}e^{-}$ pair,
and in turn the conversion efficiency drops considerably \cite{Ginz}.
In the following we take $z_{i}^{max}=0.83$.
The maximum value of the total energies of
$\gamma\gamma$ sub-processes corresponds to about
80\% of $\rs$ of the basic $ee$ colliders,
\begin{equation}
\sqrt{s_{\gamma\gamma}}<\sqrt{z_{1}^{max}z_{2}^{max}s_{ee}}\simeq
0.8\sqrt{s_{ee}}.
\label{zmax}
\end{equation}

Both $D_{\gamma/e}$ and $\xi_{2}$ are functions of
$z_{i}$ and depend on $\lambda_{i}$,
the mean helicity of electron to be scattered off by the
laser photon, and $P_{i}^c$, that of the laser photon \cite{Ginz}:
\begin{eqnarray}
D_{\gamma/e}(z_{i})&=&D_{\gamma/e}(z_{i};\lambda_{i}P_{i}^c) \nonumber \\
\xi_{2}(z_{i})&=&\xi_{2}(z_{i})(z_{i};\lambda_{i},P_{i}^c).
\end{eqnarray}
It should be mentioned that $D_{\gamma/e}(z_{i})$ depend only on
the value $\lambda_{i}P_{i}^c$.
By polarizing the $e$ beams and the laser photons, we can obtain
monochromatic and polarized gamma beams \cite{Ginz}.
For example, if we set $\lambda_{1,2}=+1/2$ and $P_{1,2}^{c}=-1$
we can get not only highly monochromatic energy spectra but also
the highly polarized initial gamma beams
with the Stokes parameters $\xi_{2}(z_{1,2})\simeq +1$ at
$z_{1,2}\simeq z_{1,2}^{max}$.

%\section{Numerical Results}

In Fig.~1 we show the slepton mass dependence of
the polarized cross sections for the sub-processes
Eq.~(\ref{total}),
where we take $\rsh$ $=$ $\sqrt{s_{\gamma\gamma}}$ $=$ 1TeV.
It is found that a choice of polarization
($\xi_{2}(z_{1})$, $\xi_{2}(z_{2})$) $=$ ($\pm$, $\pm$)
enhances the cross sections for the sleptons with masses
close to the threshold values,
because those polarized cross sections have the
$s$-wave contribution (see Eqs.(\ref{subsl}) and (\ref{total})).
We can conclude that in order to get large cross sections
for the sleptons with large masses $\msl\nle\rsh/2$,
the polarized photon beams with
($\xi_{2}(z_{1})$, $\xi_{2}(z_{2})$) $=$ ($\pm$, $\pm$)
would be efficient.
On the other hand, if the total energy $\rsh$ was large enough
$\rsh\gg 2\msl$,
($\xi_{2}(z_{1})$, $\xi_{2}(z_{2})$) $=$ ($\pm$, $\mp$)
would be useful.

Next we show the numerical results for
the experimental cross sections Eq.~(\ref{expsig}).
The slepton mass dependence of the polarized cross sections
is shown in Fig.~2, where we set $\sqrt{s_{ee}}$ $=$ 1TeV.
For comparison, we also plotted the total cross sections for
$e^{+}e^{-}\to\ser^{+}\ser^{-}$ and
$e^{+}e^{-}\to\smr^{+}\smr^{-}$,
where we take the lightest neutralino mass as
$\msz1$ $=$ $\mbino$ $=$ 50GeV.
The cross section for the
selectron production at $e^{+}e^{-}$ colliders are larger
than the $\gamma\gamma$ one, because there exist the
$t$-channel neutralino contributions.
On the other hand, the $\gamma\gamma$ cross section will be much
larger than $e^{+}e^{-}$ one for 2nd (and 3rd) generation sleptons
if $\rs$ is large compared with the mass threshold.
This is originated from the
$s$-wave contribution in Eqs.(\ref{subsl}) and (\ref{total}).
In particular, the beam polarization
$\lambda_{1,2}=+1/2$ and $P_{1,2}^{c}=-1$ will significantly
enhance the cross section.
This is because the polarization
brings us highly monochromatic energy spectra and
the highly polarized initial photon beams
($\xi_{2}(z_{1})$, $\xi_{2}(z_{2})$) $\simeq$ ($+$, $+$)
at $z_{1,2}\simeq z_{1,2}^{max}$.
Such photon polarization
enhances the sub-process cross sections
for the sleptons with the masses
$\msl$ $\nle$ $\sqrt{\sh_{max}}/2$ $\simeq$ $0.4\rs$.
Note that we can get the same enhanced cross sections
when we choose
$\lambda_{1,2}=-1/2$ and $P_{1,2}^{c}=+1$.

Now we should discuss the signatures of the processes and the background
suppression.
Here we pay attention to the right-handed sleptons $\slr$ production,
because this mass eigenstate is lighter than the left-handed one
in most of the supergravity GUTs.
It is expected
that the signature of this process will be rather simple
because $\slr$ decays dominantly into
$\ell\sz1$ \cite{decay},
where $\sz1$ denotes the lightest neutralino.
For the left-handed slepton production,
which has more complicated decay pattern,
we refer to Ref.\cite{Cuypers}.
Since we can legitimately assume $\sz1$ as the
lightest SUSY particle (LSP),
the signature of our process will be the charged lepton pair
plus large missing energies taken away by the LSP.
The most serious SM background is
$\gamma\gamma\to W^{+}W^{-} \to
(\ell^{+}\nu)(\ell^{-}{\overline{\nu}})$ \cite{Cuypers}.
We have already given the formula Eq.(\ref{subW})
for the sub-process $\gamma\gamma\to W^{+}W^{-}$ cross section.
At the TeV energy colliders, a charged lepton coming from
the $W$-boson decay will be emitted into almost same direction with
that of the parent $W$. Consequently, a cut on the acoplanarity
$\phi_{\rm acop}$ of the charged lepton pair will
be effective to suppress the
$WW$ background.

In Fig.~3 we show the Monte-Carlo event generation for the
signal and the background processes in the transverse energy
$E^{\ell}_{T}$ (a)
and the missing transverse energy $E^{\rm miss}_{T}$ (b) distributions,
where we impose a cut on the acoplanarity,
$\phi_{\rm acop}$ $>$ 40$^{\circ}$.
Here we take $\rs$ $=$ $\sqrt{s_{ee}}$ $=$ 1TeV, $\mslr$ $=$ 350GeV,
$\msz1$ $=$ 50GeV,
($\lambda_{1,2}$, $P_{1,2}^{c}$) $=$ (+1/2, -1)
and the luminosity $L$ $=$ 25fb$^{-1}$.
It is found that a lower cut on $E^{\ell}_{T}$ or $E^{\rm miss}_{T}$
will be useful for more suppression of the background.
The slepton mass dependence of the total cross section,
after imposing cuts on the missing transverse energy
$E^{\rm miss}_{T}$ $>$ 150GeV as well as on the acoplanarity
$\phi_{\rm acop}$ $>$ 40$^{\circ}$, is shown in Fig.~4,
where we take the neutralino mass as
$\msz1$ $=$ 50GeV [150GeV] in (a) [(b)].
The horizontal lines correspond to the $WW$ background cross section
after imposing the cuts.
We should be noted that the background cross section is
almost unaffected by the initial beam polarization.
We see that if we take the polarization
($\lambda_{1,2}$, $P_{1,2}^{c}$) $=$ (+1/2, -1),
the signal cross sections dominate over the background one
for the masses near the kinematical limit $\mslr\nle 0.4\rs$.
Note that this is almost independent on the neutralino mass.
On the other hand, the cross sections drops significantly for large
neutralino and slepton masses if the initial beams were not polarized.

%\section{Conclusion}
%\label{sec:conclusion}

  We have investigated the charged slepton production at
$\gamma\gamma$ colliders and
focused our attention to
the physical consequence of the initial beam polarization.
It has been shown that
appropriate beam polarization could be useful to enhance
the cross sections for the sleptons with the masses $\msl\nle 0.4\rs$.
We have explicitly shown that the most serious $WW$ background
can be suppressed by the cuts on
the acoplanarity and the missing transverse energy
as well as the choice of polarization.
For the slepton in 2nd (and 3rd) generation
the $\gamma\gamma$ cross section can be larger than
$e^{+}e^{-}$ one if we take appropriate initial beam polarization and
the large total energies $\rs\nge 2.5\msl$.
This property is originated from the
$s$-wave contribution in the scalar pair production cross section.
This could enable us to see, moreover, the squarkonium
(especially the stoponium) production
in the $\gamma\gamma$ colliders, which could not be seen
in the $e^{+}e^{-}$ colliders \cite{proc,prep}.
Another good property of the processes is the simple
dependence on the arbitrary SUSY parameters, i.e.,
the cross sections depend only on the final sfermion masses.
This could be useful to check the universality of masses
of sleptons and squarks in the 1st and 2nd generation.

\vskip 20pt
\begin{flushleft}
{\Large{\bf Acknowledgement}}
\end{flushleft}
The author thanks A. Goto for his collaboration at early stage
of this work.

\vfill\eject
\baselineskip = 18pt plus 1pt
\noindent{\Large{\bf Figure Captions}} \\
\medskip
\nobreak
{\bf Figure 1:} \ \
Scalar lepton mass dependence of total cross sections
for each photon polarization
($\xi_{2}(z_{1})$, $\xi_{2}(z_{2})$).
We take $\sqrt{s_{\gamma\gamma}}$ $=$ 1TeV. \\
\medskip
{\bf Figure 2:} \ \
Scalar lepton mass dependence of total cross sections
for each initial beam polarization
($\lambda_{1}$, $P_{1}^{c}$) and ($\lambda_{2}$, $P_{2}^{c}$).
We take $\sqrt{s_{ee}}$ $=$ 1TeV. \\
\medskip
{\bf Figure 3:} \ \
Monte-Carlo event generation
for $\gamma\gamma\to\sl\slb$ and $\gamma\gamma\to W^{+}W^{-}$
with cut $\phi_{\rm acop}$ $>$ 40$^{\circ}$.
We take $\sqrt{s_{ee}}$ $=$ 1TeV, $\mslr$ $=$ 350GeV,
$\msz1$ $=$ 50GeV,
($\lambda_{1,2}$, $P_{1,2}^{c}$) $=$ (+1/2, -1)
and the luminosity $L$ $=$ 25fb$^{-1}$. \\
{\bf Figure 4:} \ \
Scalar lepton mass dependence of total cross sections
with cuts $E^{\rm miss}_{T}$ $>$ 150GeV and
$\phi_{\rm acop}$ $>$ 40$^{\circ}$.
We take $\sqrt{s_{ee}}$ $=$ 1TeV and
$\msz1$ $=$ 50GeV for (a) and 150GeV for (b).
Horizontal lines correspond to $WW$ background cross section
after imposing the cuts.
\medskip
\goodbreak
\bigskip

\vskip 20pt


\begin{thebibliography}{99}
\bibitem{Ginz}
 I. F. Ginzburg, G. L. Kotkin, V. G. Serbo and V. I. Telnov,
{\it Nucl. Instrum. Methods}, {\bf 205} (1983) 47 ;
I. F. Ginzburg, G. L. Kotkin, S. L. Panfil, V. G. Serbo and V. I. Ternov,
{\it Nucl. Instrum. Methods}, {\bf 219} (1984) 5
\bibitem{Brodsky}
S. Brodsky, Plenary talk at
{\it the 2nd International Workshop on Physics
and Experiments at Linear $e^{+}e^{-}$ Colliders},
(Waikoloa, Hawaii, 26-30, April, 1993)
\bibitem{Borden}
D. Borden, Plenary talk at
{\it the 2nd International Workshop on Physics
and Experiments at Linear $e^{+}e^{-}$ Colliders},
(Waikoloa, Hawaii, 26-30, April, 1993)
\bibitem{gauge}
 I. F. Ginzburg, G. L. Kotkin, S. L. Panfil and V. G. Serbo,
{\it Nucl. Phys.} {\bf B228} (1983) 285 ;
A. Grau and J. A. Grifols, {\it Nucl. Phys.} {\bf B233} (1984) 375 ;
P. Kalyniak, G. Couture and S. Godfrey, Proc. of the
\uppercase\expandafter{\romannumeral12}
Warsaw Symposium on Elementaly Particle Physics, ed. by Z. Ajduk et al.,
(World Scientific, 1990) p.595 ;
S. Y. Choi and F. Schrempp, {\it Phys. Lett.} {\bf B272} (1991) 149 ;
{\it Proc. of the Workshop on $e^{+}e^{-}$ Collisions at 500GeV :
The Physics Potential}, ed. by P. M. Zerwas, DESY 92-123B (1992) p.793 ;
K. Cheung, preprint, NUHEP-TH-92-24 (1992) ;
G. B\'elanger and F. Boudjema,
{\it Proc. of the Workshop on $e^{+}e^{-}$ Collisions at 500GeV :
The Physics Potential}, ed. by P. M. Zerwas, DESY 92-123B (1992) p.783
\bibitem{Higgs}
 R. Najima, Proc. of the 3rd Meeting on Physics at TeV Energy Scale,
ed. by K. Hidaka et al., KEK Report 90-9 (1990) p.112 ;
I. F. Ginzburg, Novosibirsk preprint, 28 (182) (1990) ;
E. E. Boos and G. V. Jikia, {\it Phys. Lett.} {\bf B275} (1992) 164 ;
 E. Boos, M. Dubinin, V. Ilyin, A. Pukhov, G. Jikia and S. Sultanov,
{\it Phys. Lett.} {\bf B273} (1991) 173 ;
 K. Hagiwara, I. Watanabe and P. M. Zerwas,
{\it Phys.Lett.} {\bf B278} (1992) 187 ;
F Richard,
{\it Proc. of the Workshop on $e^{+}e^{-}$ Collisions at 500GeV :
The Physics Potential}, ed. by P. M. Zerwas, DESY 92-123B (1992) p.883 ;
K. Cheung, preprints, NUHEP-TH-92-21 (1992) ; NUHEP-TH-93-3 (1993) ;
D. Bowser-Chao and K. Cheung, preprint, NUHEP-TH-92-29 (1993) ;
K Melnikov and O. Yakovlev, preprint, BUDKERINP 93-4 (1993)
\bibitem{QCD}
M. Drees and R. M. Godbole,
{\it Proc. of the Workshop on $e^{+}e^{-}$ Collisions at 500GeV :
The Physics Potential}, ed. by P. M. Zerwas, DESY 92-123B (1992) p.863 ;
 O. J. P. \'Eboli et al.,
{\it Phys.Lett.} {\bf B301} (1993) 115
\bibitem{egamma}
 A. Goto and T. Kon, {\it Europhys. Lett.}
{\bf 19} (1992) 575 ;
T. Kon and A. Goto, {\it Phys. Lett.} {\bf B295} (1992) 324
H. K\"{o}nig and K. A. Peterson, {\it Phys. Lett.}
{\bf B294} (1992) 110 ;
F. Cuypers, G. J. van Oldenborgh and R. R\"{u}ckl,
{\it Nucl. Phys.} {\bf B383} (1992) 45 ; preprint
CERN-TH-6742/92 (1992)
\bibitem{gamma}
 A. Goto and T. Kon, {\it Europhys. Lett.}
{\bf 13} (1990) 211 ;
{\bf 14} (1991) 281 ;
\bibitem{Cuypers}
F. Cuypers, G. J. van Oldenborgh and R. R\"{u}ckl, preprint
CERN-TH-6742/92 (1992)
\bibitem{proc}
 T. Kon, Talks at
{\it the 4th Workshop on JLC}, (KEK, 15--16, March, 1993)
and {\it the 2nd International Workshop on Physics
and Experiments at Linear $e^{+}e^{-}$ Colliders},
(Waikoloa, Hawaii, 26-30, April, 1993) ;
Seikei Univ. preprint, ITP-SU-93/02 (1993)
\bibitem{GI}
 I. F. Ginzburg and D. Yu. Ivanov, {\it Phys. Lett.} {\bf B276} (1992) 214
\bibitem{egest}
 T. Kon, I. Ito and Y. Chikashige, {\it Phys. Lett.} {\bf B287} (1992) 277
\bibitem{Eboli}
 O. J. P. \'Eboli et al., preprint, IFT-P.014/93
\bibitem{decay}
 A. Bartl, W. Majerotto and B. M\"osslacher,
{\it Proc. of the Workshop on $e^{+}e^{-}$ Collisions at 500GeV :
The Physics Potential}, ed. by P. M. Zerwas, DESY 92-123B (1992) p.641
\bibitem{prep}
 T. Kon, in preparation
\end{thebibliography}
\end{document}